\documentclass[runningheads]{llncs}
\usepackage{graphicx}
\begin{document}
\title{Left atrial ejection fraction estimation using SEGANet for fully automated segmentation of CINE MRI } 
\titlerunning{SEGANet: Automated segmentation of the left atrium in CINE-MRI}
%
\author{Ana Lourenço\inst{1,2}
,Eric Kerfoot\inst{1}
Connor Dibblin\inst{1}
,Ebraham Alskaf\inst{1}
,Mustafa Anjari\inst{3}
,Anil A Bharath\inst{4}
,Andrew P King\inst{1}
,Henry Chubb\inst{1}
,Teresa M Correia\thanks{Contributed equally.}\inst{1}
,Marta Varela*\inst{1,5}
}
%
\authorrunning{A. Lourenço et al.} 

%
\institute{School of Biomedical Engineering and Imaging Sciences, King's College London, London, United Kingdom \and
Faculty of Sciences, University of Lisbon, Lisbon, Portugal \and  Department of Neuroradiology, National Hospital for Neurology and Neurosurgery, University College London Hospitals NHS  Trust, London, United Kingdom \and Bioengineering Department, Imperial College London, London, United Kingdom \and National Heart and Lung Institute, Imperial College London, London, United Kingdom}

\maketitle              
\begin{abstract}
Atrial fibrillation (AF) is the most common sustained cardiac arrhythmia, characterised by a rapid and irregular electrical activation of the atria. 
Treatments for AF are often ineffective and few atrial biomarkers exist to automatically characterise atrial function and aid in treatment selection for AF. Clinical metrics of left atrial (LA) function, such as ejection fraction (EF) and active atrial contraction ejection fraction (aEF), are promising, but have until now typically relied on volume estimations extrapolated from single-slice images. \\
In this work, we study volumetric functional biomarkers of the LA using a fully automatic SEGmentation of the left Atrium based on a convolutional neural Network (SEGANet). SEGANet was trained using a dedicated data augmentation scheme to segment the LA, across all cardiac phases, in short axis dynamic (CINE) Magnetic Resonance Images (MRI) acquired with full cardiac coverage. Using the automatic segmentations, we plotted volumetric time curves for the LA and estimated LA EF and aEF automatically.  \\
The proposed method yields high quality segmentations that compare well with manual segmentations (Dice scores [$0.93 \pm 0.04$], median contour [$0.75 \pm 0.31$] mm and Hausdorff distances [$4.59 \pm 2.06$] mm). LA EF and aEF are also in agreement with literature values and are significantly higher in AF patients than in healthy volunteers. Our work opens up the possibility of automatically estimating LA volumes and functional biomarkers from multi-slice CINE MRI, bypassing the limitations of current single-slice methods and improving the characterisation of atrial function in AF patients.
\end{abstract}
%
%
\section{Introduction}
Atrial fibrillation (AF) is the most common sustained cardiac arrhythmia, characterized by a rapid and irregular contraction of the upper chambers of the heart, the atria. AF affects more than 33 million people worldwide, is associated with great increases in morbidity and mortality, and typically has a negative impact on patients' quality of life.
Although AF is mainly managed with medical therapy, catheter ablation is arguably the gold standard treatment to directly terminate AF in selected patients. In their current form, ablations are only first-time effective in approximately $50\%$ of AF patients \cite{Ganesan2013}.

Treatment efficacy for AF could be greatly increased if biomarkers capable of characterising atrial function and predicting treatment outcome were available. Left atrial (LA) dimensions and shape have been shown to be reasonable predictors of post-ablation patient outcome 
\cite{Varela2017}. It is expected that dynamic  characterisations of atrial dimensions and volumes across the cardiac cycle may be even more informative \cite{Hoit2017}. In particular, LA ejection fraction (EF) and active atrial contraction ejection fraction (aEF) contain important clinical information that can contribute to the management of AF.
The LA EF characterises the LA global function across the cardiac cycle, whereas the aEF assesses the LA pump function, which is likely to provide important additional clues about atrial tissue health. Both these biomarkers rely on ratios of LA volumes across the cardiac cycle. LA volumes are usually estimated by applying standard formulae derived from assumptions on the LA shape obtained from 2D atrial echocardiography or dynamic (CINE) Magnetic Resonance Images (MRI) \cite{ERBEL,UJINO20061185}. Volume estimation from 3D images is expected to be more accurate and precise. Furthermore, aEF relies on the identification of the onset of active atrial contraction (preA), which is difficult to perform visually in a single 2D view.  

CINE MRI combines good spatial and temporal resolution with large cardiac coverage and is thus ideally suited for assessing LA function. Nevertheless, detailed volumetric temporal analyses of atrial volumes are not currently available for clinical use for two key reasons. First, atrial CINE MRI images are typically acquired on single-slice 2-chamber and 4-chamber views, which do not sample the full atrial volume \cite{Kowallick2015}. Second, reliable automatic techniques for segmenting the atria in short axis views across the entire cardiac cycle are not yet available. This is in contrast with the segmentation of ventricular structures from short axis CINE MRI and the LA in 2- and 4-chamber views~\cite{Chen2019}. We expect that automatic processing tools for short axis atrial images may lead to an increased interest in imaging this chamber in this orientation.

In this work, we propose a dedicated neural network for fully automatic SEGmentation of the left Atrium based on a convolutional neural Network (SEGANet) to address this important clinical gap. We acquired short-axis CINE images with full ventricular and atrial coverage and used SEGANet to segment the LA. The quality of the SEGANet LA segmentation is assessed using the Dice coefficient (DC), median contour distances (MCD) and Hausdorff distances (HD). We additionally compare the SEGANet results with the inter-observer variability of manual segmentations. To demonstrate the clinical utility of our method, we use SEGANet to estimate LA volumes across all cardiac phases and automatically calculate LA EFs and aEFs in both healthy subjects and AF patients.

\section{Methods}
\subsection{Data Acquisition}

Short axis CINE MR image stacks with full cardiac coverage were acquired in a 1.5T Philips Ingenia scanner with a 32-channel cardiac coil. Imaging was performed in 60 AF patients (31-72 years old, $75 \%$ male) and 12 healthy volunteers (24-36 years old, $50 \%$ female) under ethical approval and following informed written consent. 

A 2D bSSFP protocol (flip angle: $60^{\circ}$, TE/TR: 1.5/2.9 ms, SENSE factor 2) was used. Images were captured with ECG-based retrospective gating with a typical field of view of 385 x 310 x 150 mm$^3$, acquisition matrix of 172 x 140 and slice thickness of 10 mm. Images were reconstructed to a resolution of 1.25 x 1.25 x 10 mm$^3$ and 30-50 cardiac phases with $60-70\%$ view sharing. These acquisitions provide full coverage of the ventricles and atria, with typically 4-6 slices covering the atria. Each slice was acquired in a separate breath hold. 
Due to the large slice thickness in all datasets, it can be difficult, even for an expert, to identify the separation between the atria and ventricles.
In order to solve this challenge, we used a pre-trained left ventricular (LV) segmentation network \cite{Kerfoot2018} to identify the slices containing ventricular tissue.
The most basal atrial slice was automatically identified as the slice contiguous to the top ventricular slice segmented by this LV network.

Ground truth (GT or M1) segmentations were obtained by a consensus of two medical experts, who manually segmented the LA on a slice by slice basis at 3 phases of the cardiac cycle. 
The LA appendage was included in the segmentation, but the pulmonary vein insertions were not.


\subsection{SEGANet}
\begin{figure}[t]
{\includegraphics[width=0.75\linewidth]{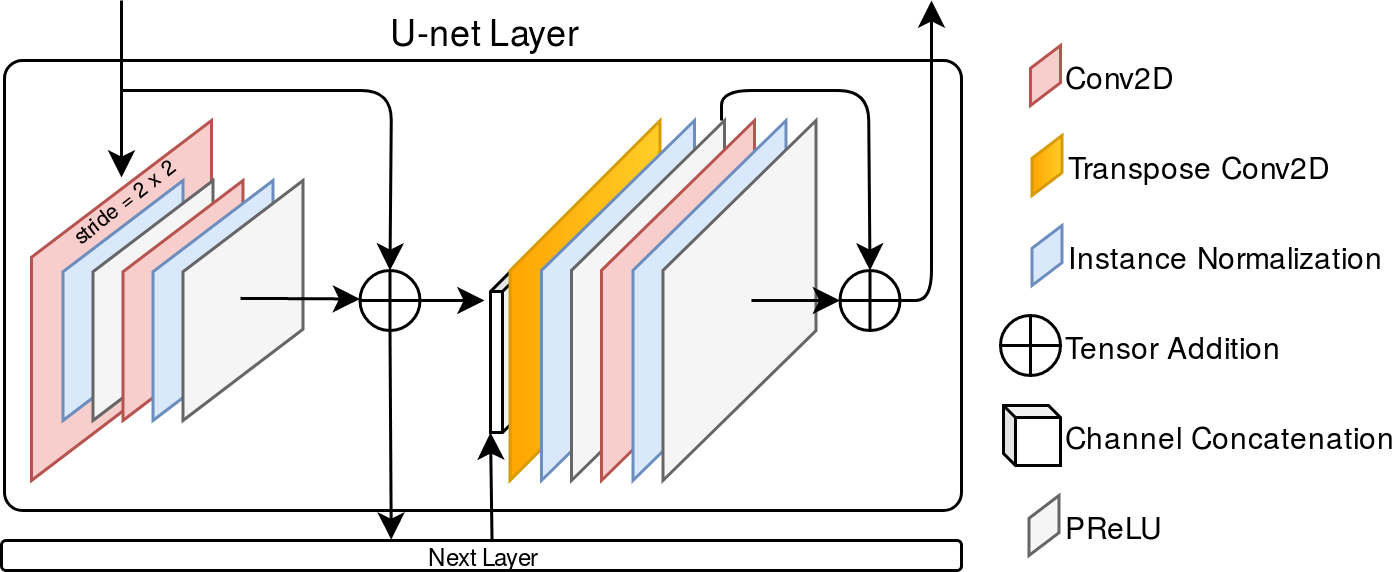}\includegraphics[width=0.18\linewidth]{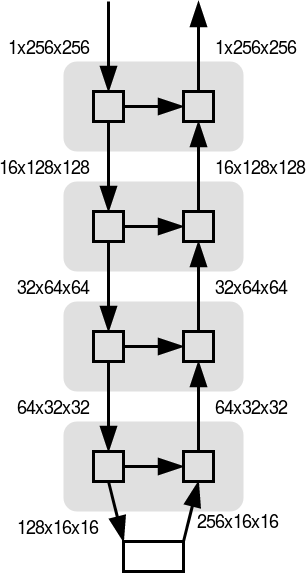}}
\caption{{The proposed LA segmentation network SEGANet is based on a U-Net architecture. The segmentation network is built as a stack of layers (shown on the right), which combine the encode path of the network on the left and the decode path on the right. Subsequent layers, or the bottom connection, are denoted by ``Next Layer''. The bottom connection is composed of a single residual unit including two sets of convolution/normalisation/regularisation sequences.}} \label{fig:unet}
\end{figure}

The proposed neural network is based on the U-Net architecture~\cite{DBLP:journals/corr/RonnebergerFB15} with the following modifications: 1) the addition of residual units~\cite{DBLP:journals/corr/HeZR016} throughout the encode and decode paths, which prevent vanishing gradients by allowing some measure of the original input to be included in each layer’s output; 2) instance  layer-normalization, which minimises the influence outlier images may have in the whole batch; 3) Parametric  Rectified  Linear  Unit  (PReLU)  activation, which allows the network to learn from negative values. 
The network is built as a stack of layers which combine the encode and decode paths as illustrated in Figure \ref{fig:unet}. Four layers are stacked together using convolutions with a stride of 2 to downsample data in the encode path, and with output tensors to the layers below with channel dimensions of 16, 32, 64, 128, and 256.

The network was trained for LA segmentation on short axis CINE MRI in $50,000$ iterations with mini-batches of 300 slices drawn randomly from the 715 slices of the training dataset. 
We used a binary formulation of the Dice loss function 
and the Adam optimiser 
with a learning rate of 0.0001.

To extract more generalised information from our relatively small dataset and to regularise the network to prevent overfitting, we used a number of randomly-chosen data augmentation functions to apply to each slice. These data augmentations were chosen to replicate some of the expected variation in real images without modifying the cardiac features of interest. They include: 1) rotations, translations and flips; 2) croppings and rotations; 3) additive stochastic noise; 4) k-space corruption to simulate acquisition artefacts;
5) smooth non-rigid deformations encoded using free-form deformations
and 6) intensity scalings by a random factor.


\subsection{Data Analysis}
In order to assess the quality of the automatic segmentations, an additional medical expert was asked to provide manual LA segmentations (M2) in 13 AF patients. Segmentation quality metrics (DC, HD and MCD) were estimated relative to the ground truth for both the SEGANet segmentations and the manual segmentations M2. 
We also estimated the agreement between the two sets of manual segmentations.

For a complete visualisation of the heart volume variations throughout the cardiac cycle, LA volumetric time curves were plotted. Using these, maximal ($V_{max}$) and minimal ($V_{min}$) LA volumes, as well as the LA volume at the onset of atrial contraction ($V_{pre A}$) were automatically computed. EF and aEF were also estimated using $ \textrm{EF} = \frac{V_{max}-V_{min}}{V_{max}} \times 100$ and $ \textrm{aEF} = \frac{V_{pre A}-V_{min}}{V_{pre A}} \times 100$.

LA EFs and aEFS obtained from SEGANet segmentations were calculated for both AF patients in sinus rhythm and healthy subjects. We used paired Student t-tests to assess whether we could detect a significant difference in these metrics between the two groups.

\begin{figure}[t]
\begin{center}
\includegraphics[width=1\linewidth, height=3.5cm]{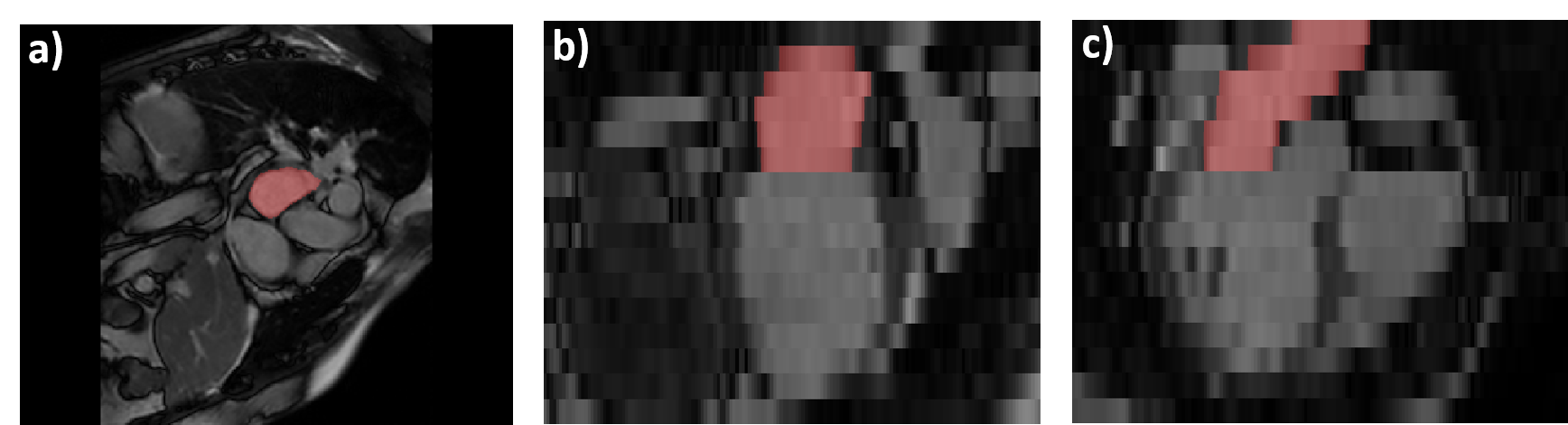} 
\caption{Segmentations obtained with the proposed method for one representative subject during atrial diastole, overlaid on the CINE  \textbf{a)} short axis, \textbf{b)} pseudo 2-chamber, and \textbf{c)} pseudo 4-chamber views. The acquired short axis CINE image stacks include both ventricles and atria.} 
\label{fig:3D}
\end{center}
\end{figure}
\section{Results}
\begin{figure}[t]
    \centering
         \includegraphics[width=1\textwidth]{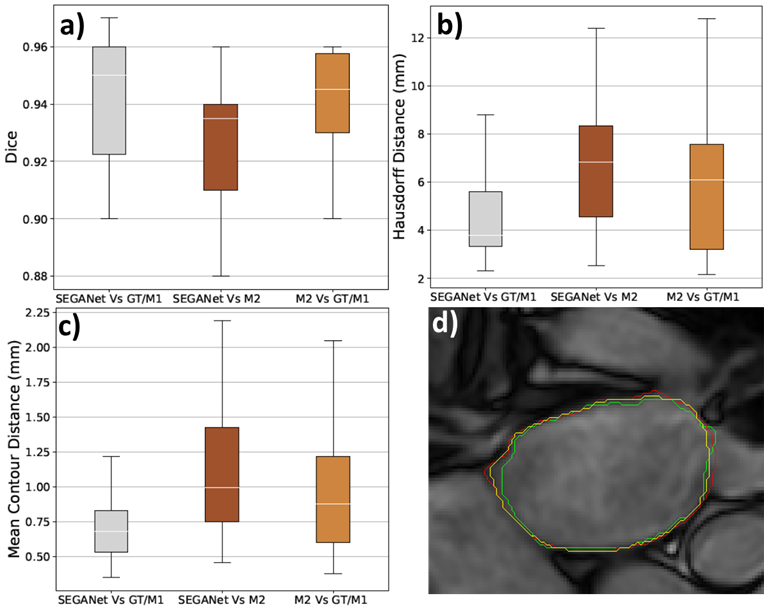}
\caption{ \textbf{a)} Dice coefficient, \textbf{b)} Hausdorff distance and \textbf{c)} mean contour distance comparing automatic (SEGANet) and manual (M2) LA segmentations with the ground truth (GT/M1) in 13 subjects. \textbf{d)} Contours indicating the results of the manual and automatic LA segmentations in a representative slice: yellow is the SEGANet automatic segmentation, red is the manual M2 segmentation, and green is the GT/M1.}
\label{fig:abs2}
\end{figure}

\hspace{1em} SEGANet generated very good quality LA segmentations. This can be qualitatively observed in Figure \ref{fig:3D}, which depicts, in three orthogonal views, the CINE MRI for one representative patient overlaid with the automatic LA segmentations in red. 
Figure \ref{fig:abs2} shows that automatic segmentations are on par with manual segmentations, with DC: $0.93 \pm 0.04$, HD: $4.59 \pm 2.06$ mm; MCD: $0.75 \pm 0.31$ mm for the SEGANet segmentations when compared to the GT. These values are comparable to those of the additional manual segmentations M2, further highlighting the good performance of the network (Figure \ref{fig:abs2}).

In previous studies, the network for LV segmentation had demonstrated accurate results when it was tested using other datasets  \cite{Kerfoot2018} and the method allowed us to automatically select the slices that covered the atria.

\begin{figure}[h]
    \centering
         \includegraphics[width=1\textwidth]{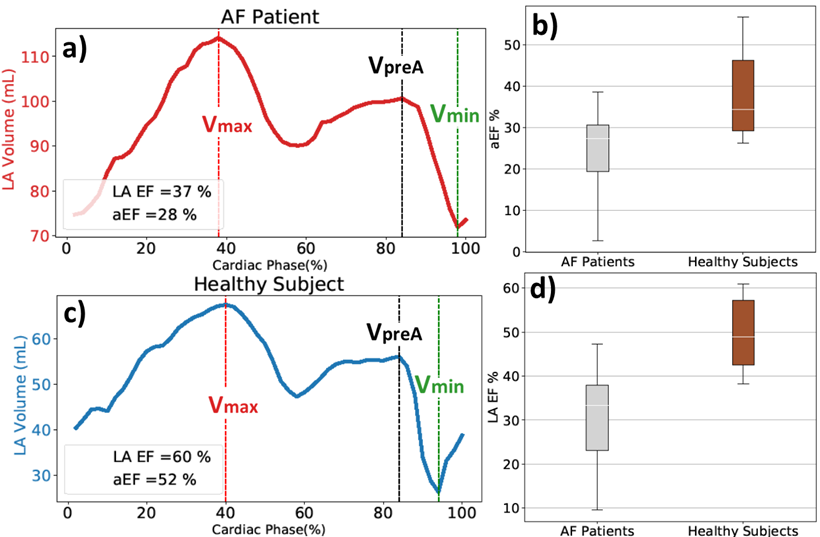}
\caption{Left atrium (LA) volume throughout the cardiac cycle for two representative subjects: \textbf{a)} an AF patient and \textbf{c)} an healthy subject . Three LA volumes are indicated: the maximal volume ($V_{max}$), the minimal volume ($V_{min}$) and the second peak volume, corresponding to the volume at the onset of the P wave, just before active atrial contraction ($V_{pre A}$). LA EF and aEF values are shown in each subfigure. Moreover, box plots of \textbf{b)} aEF and \textbf{d)} LA EF for AF patients and healthy subjects show that values are significantly higher in healthy subjects than in patients (LA EF p-value: $<1\textrm{e}^{-4}$; aEF p-value $<1\textrm{e}^{-7}$).}
\label{fig:h_vs_p}
\end{figure}

The obtained LA volumes across the cardiac cycle for AF patients ($V_{min}: 79.40 \pm 25.34$ mL; $ V_{max}: 111 \pm 24.00$ mL; $ V_{pre A}: 103.40 \pm 25.34$ mL) and for healthy subjects ($V_{min}: 22.43 \pm 25.34$ mL; $ V_{max}: 44.23 \pm 24.73$ mL; $ V_{pre A}: 35.47 \pm 19.48 $ mL) show high inter-subject variability and agree well with the literature \cite{MORI201189}.
 The LA volume varies smoothly across the cardiac cycle, further suggesting the good quality of segmentations throughout the different cardiac phases (Figure \ref{fig:h_vs_p}).    

Regarding the LA EF, we obtained $31.1\%  \pm 9.9 \%$ in AF patients and $49.8\% \pm 7.6 \%$ in healthy volunteers, which is in good agreement with published echocardiographic and 2D CINE MRI values \cite{Chubb2018,MORI201189,DELGADO20081285}. 
The obtained values for aEF (AF patients: $24.3\%  \pm 9.0 \%$, healthy volunteers: $37.9\%  \pm 10.1 \%$) are also in accordance with the literature
\cite{Kowallick2015,MORI201189,DELGADO20081285}. 
As expected, both LA EF and aEF are significantly higher ($p<1\textrm{e}^{-4}$ and $p<1\textrm{e}^{-7}$, respectively) in healthy subjects than in AF patients (Figure \ref{fig:h_vs_p}). 

\section{Discussion}
\hspace{1em}
We present SEGANet, a 
convolutional neural network that allows automatic LA segmentation in short-axis CINE MRI across the whole cardiac cycle. To the best of our knowledge, there are currently no 
fully automatic segmentation methods designed for this important clinical application. We demonstrate the clinical potential of our technique by automatically estimating atrial biomarkers of interest, such as atrial ejection fractions.

The proposed method shows a very good segmentation performance for the LA (Figures \ref{fig:3D}-\ref{fig:abs2}), on par with the performance of other neural networks: for the segmentation of ventricular structures in CINE-MRI \cite{Chen2019}; 
and for LA segmentations in static MRI with a different resolution and contrast to ours. In particular, our results for CINE MRI compare well with the metrics for the segmentation of delayed enhacement MRI in the STACOM 2018 Left Atrial Segmentation Challenge (DC: $0.90 \pm 0.93$, HD: $14.23 \pm 4.83$ mm)\cite{inbook} .


In this work, we rely on a modified short axis CINE MRI acquisition that provides volumetric and functional information for the LA, as well as for the ventricles.
To keep our analysis pipeline fully automatic, we employed a pre-trained LV segmentation network \cite{Kerfoot2018} to identify the most basal atrial slice. This approach was found to be extremely reliable for all images and carried a negligible time penalty.

We chose to segment multi-slice CINE MRI data on a slice-by-slice basis. Although this setup does not make direct use of the 3D spatial and temporal correlations of this dataset, this approach has other advantages, as highlighted previously \cite{Chen2019}. 
One of the advantages is the comparative simplicity and low parameter numbers of the SEGANet compared to its 3D or 3D+time counterparts and its immunity to misregistrations across slices caused by variations in breath hold positions. Additionally, treating each slice independently allows us to train the network with comparatively fewer number of patient datasets than for 3D (or 3D+time) networks. Furthermore, due to the relatively small size of the training dataset (715 slices), we applied an extensive data augmentation strategy, which was crucial for the good performance shown by the network. We add that our segmentation also uniquely includes the LA appendage, which is relevant for assessing stroke risk in AF. 

To demonstrate the clinical utility of our work, we used SEGANet to obtain fully automatic estimates of clinical biomarkers such as LA volumes, LA EFs and aEFs. Up until now, LA volumes and functional markers have been estimated in CINE MRI by applying empirical formulae to single-slice long axis views \cite{ERBEL,UJINO20061185}, after manual annotation. These shortcomings have contributed to a comparative lack of interest in imaging left atrial function using CINE MRI. 
Our method allows instead for biomarker estimates from volumetric CINE MRI, which are expected to be more accurate, reproducible and thus more clinically useful.

Estimates of LA volumes from long axis 2D slices typically rely on manual delineations of left atrial apex-base dimensions and assume, unrealistically (see Fig. \ref{fig:3D}), that the LA shape is as geometrically regular as that of the LV. Furthermore, given that atrial deformations across the cardiac cycle are not isotropic, it is not clear whether single 2D views allow for the accurate identification of the cardiac phases corresponding to maximal LA and minimal LA volume, as required for EF estimation. It is particularly difficult to identify the onset of atrial contraction ($V_{preA}$), required for aEF estimation, which our method performs automatically. 

We thus expect that SEGANet will improve the clinical usefulness of atrial CINE MRI imaging and ultimately 
make an important contribution to the stratification and treatment of AF patients.



\section{Conclusion}
\hspace{1em}
We have presented SEGANet for fully automatic segmentation of the LA from CINE MRI and showed it can accurately segment the LA across the cardiac cycle. SEGANet provides reliable automatic estimates of clinical biomarkers such as LA volumes, LA EF and aEF. 


\vspace{2em}
\noindent \textbf{\large{Acknowledgments}}
\vspace{2em}

This research was supported by the Wellcome EPSRC Centre for Medical Engineering at the School of Biomedical Engineering and Imaging Sciences, King’s College London [WT203148/Z/16/Z] and the National Institute for Health Research (NIHR) Biomedical Research Centre at Guy's and St Thomas' NHS Foundation Trust and King's College London. We also acknowledge funding from the Engineering and Physical Sciences Research Council [EP/N026993/1] and the British Heart Foundation [RE/18/4/34215].).

\bibliography{main_1}
\bibliographystyle{splncs04}

\end{document}